\documentclass[useAMS,usenatbib]{mn2e}
\usepackage{amsmath}
\usepackage{amsfonts}
\usepackage{amssymb}
\usepackage{psfig}
\usepackage{color}   

\title[The Logarithmic Potential]
  {A Study of the Orbits of the Logarithmic Potential for Galaxies}
\author[S.R. Valluri et al.]
  {S.~R.~Valluri,$^{1,2,3}$\thanks{Email: valluri@uwo.ca}
  P.~A.~Wiegert,$^1$ J. Drozd,$^4$, M. Da Silva,$^1$  \\
 $^1$Department of Physics and Astronomy, The University of Western Ontario, London Ontario N6A 3K7 Canada \\
 $^2$Department of Applied Mathematics, The University of Western Ontario, London Ontario N6A 5B7 Canada \\
 $^3$King's University College, London Ontario N6A 2M3 Canada \\
 $^4$Robarts Research Institute, London Ontario N6A 5K8 Canada }
\date{Accepted 6 Sept 2012}

\pagerange{\pageref{firstpage}--\pageref{lastpage}} \pubyear{2012}

\def\LaTeX{L\kern-.36em\raise.3ex\hbox{a}\kern-.15em
    T\kern-.1667em\lower.7ex\hbox{E}\kern-.125emX}

\begin{document}

\label{firstpage}

\maketitle

\begin{abstract}	

The logarithmic potential is of great interest and relevance in the study of the
dynamics of galaxies. Some small corrections to the work of
\cite{consei90} who used the method of \cite{pre82} to find periodic
 orbits and bifurcations within such a potential are presented. The
solution of the orbital radial equation for the purely radial
logarithmic potential is then considered using the p-ellipse
(precessing ellipse) method pioneered by \cite{str06}. This
differential orbital equation is a special case of the generalized
Burgers equation. The apsidal angle is also determined, both
numerically as well as analytically by means of the Lambert $W$ and
the Polylogarithm functions. The use of these functions in computing
the gravitational lensing produced by logarithmic potentials is
discussed.
\end{abstract}


\section{Introduction}

The logarithmic potential has great interest in connection with the
dynamics of elliptical galaxies and galactic halos. Introduced by
\cite{ric80} to model stellar systems with concentric axisymmetric
oblate spheroidal potential surfaces, it is one of the few
axisymmetric galactic potentials with an equally simple mass
distribution function. As a result it has been studied extensively
({\it e.g.} \cite{binspe82,bintre87}).

\cite{ric82} did an extensive survey of orbits within scale-free
logarithmic potentials (that is, with zero core radii). The effect
of core radius and the presence or absence of a central mass have been
examined by \cite{gerbin85,pfedez89,mirsch89}.  \cite{eva93} examined the axisymmetric
case of galaxies embedded in extended dark matter
halos. \cite{leesch92} examined triaxial halo models.
\cite{karcar01} examined how the core radius and the angular momentum
are related to transitions from regular motion to chaos in log
potentials.  \cite{toutre97} developed a symplectic map to study the
dynamics of orbits in non-spherical potentials, with particular
emphasis on the logarithmic potential.  Periodic orbits in triaxial
logarithmic potentials have been examined analytically 
\cite[]{belbocpuc07,pucbocbel08} and numerically \cite[]{mag82}. 

Beyond galactic dynamics, the potential also has applications to the
problem of gravitational lensing. Beyond astrophysics, applications of
the logarithmic potential occur in the solution of planar boundary
value problems in potential theory \cite[]{eva27} and with boundary
value problems of analytic function theory. In elementary particle
physics, \cite{quiros77} use the logarithmic potential to show that
the quarkonium level spacings are independent of quark mass, in the
non-relativistic limit. In this paper, the analysis of \cite{consei90}
(CS) is re-examined. CS applied the analytical techniques
of \cite{pre82} to find approximate solutions to the equations of
motion for particles moving within a logarithmic potential.  The
Prendergast method was introduced to approximate some complex
differential equations, such as the Duffing equation, and new
applications for this method are still being found today. We then
elaborate on the work of CS, turning our attention to the radial
orbital equation, using the nonlinear Burgers equation to determine an
approximate analytic solution from which the apsidal angle is
determined. It is also determined by finding the roots of the Lambert
$W$ and the polylogarithmic function.

In sections \ref{upsection} and \ref{psection}, the Prendergast Method
is revisited. We performed a thorough study of the pioneering work of
\cite{consei90} and present a slight elaboration as well as a few
minor corrections to their equations.  In section \ref{pellsection},
we briefly study \cite{str06}'s p-ellipse (precessing ellipse),
introduced in his fine work on precessing orbits in a variety of power-law
potentials, some shallower than the $1/r$ Keplerian one. These potentials
include the logarithmic potential of zero as well as nonzero core
softening length. We present an integrable equation that provides us
with values for the apsidal angles of the orbits considered. In
section~\ref{gravlensing} we discuss the deflection of light in a
logarithmic potential and gravitational lensing. Finally,
section \ref{conclusions} summarizes our conclusions and any further
work to be considered.

\section{Unperturbed Solutions} \label{upsection}

Here we begin by re-establishing the results of CS with some minor
corrections, before going on to use these solutions in subsequent
sections. Where alterations to their values are given, they are
indicated by asterisks.

Following the notation of CS, our expression for the logarithmic potential is
\begin{equation}
V(x,y)=\ln \left( x^2+\frac{y^2}{U^2} +C^2 \right) \label{eq:logpotl}
\end{equation}
where $C$ is the core radius and $U$ describes the ellipticity of the
potential. Though of mathematical interest over a wider range of
parameters, models with $U> 1.08$ or $U < 1/\sqrt{2} = 0.707$ are
unphysical in that they require negative mass densities \cite[]{eva93}.
As a result, only values of $0.707 < U < 1.08$ are of interest to
galactic dynamics. The CS method begins by finding a solution for
arbitrary values of $U$ in the one-dimensional case ($y \equiv \dot{y}
\equiv 0$), and adding the motion in the second dimension as a
perturbation.

By finding the derivative of Eq.~\ref{eq:logpotl} and introducing it in the
relevant second order orbital differential equation, it is possible to
develop two equations of motion--one for the $x$ component, the other
for the $y$ component:

\begin{equation} 
x''+ \frac{2U^2x}{U^2x^2 + y^2 + C^2U^2}=0 \label{eq:orbitalx}
\end{equation}
\begin{equation}
y''+\frac{2y}{U^2x^2 + y^2 + C^2U^2}=0\hspace{0.75cm} (*)
\end{equation}
where the $(*)$ indicates the equation contains a correction to a typo in CS's original.

The subsequent solution is developed using the method
of \cite{pre82}. Developed for second-order nonlinear ordinary
differential equations, Prendergast applied the technique to the van
der Pol oscillator and Duffing's equation. CS applied it to the
orbital equation in the logarithmic potential.

The method begins by assuming a solution for $x$ and $y$ of the
following form:
\begin{equation}
x=\frac{N}{D}; \indent  \label{eq:xasfrac}
y=\frac{M}{D}; \indent  
\end{equation}
where $N$, $M$ and $D$ are Fourier series of the form
\begin{eqnarray}
N &=& \sum_{k={\rm odd}} N_k \cos(k \omega t), \nonumber \\
D &=& 1 + \sum_{l={\rm even}} D_l \cos(l \omega t) \label{eq:propsoln} 
\end{eqnarray}
and which are truncated at the appropriate order.  In the unperturbed
one-dimensional case, $y = \dot{y} = 0$ and $M=0$.

In determining the solution for $x$, we introduce the expansion
of \ref{eq:xasfrac} into Eq.~\ref{eq:orbitalx}, and solve for a
new equation of motion,
\begin{eqnarray}
(N''D^2-2N'D'D-ND''D+2ND'^2)\times \nonumber \\ (U^2N^2+M^2+C^2U^2D^2) 
+ 2U^2ND^4=0. \label{eq:eqnofmot}
\end{eqnarray}
A solution is essayed of the form
\begin{equation}
N=A\cos (\omega t); \indent D=1+B\cos (2\omega t); \label{eq:AandB}
\end{equation}
with constants $A$, $B$ and $\omega$ to be determined, though we require
$B\neq 0$ for a non-trivial rational approximation.

Finally, we introduce the proposed solutions \ref{eq:AandB} into
Eq.~\ref{eq:eqnofmot} and set equal to zero the coefficients of
$\cos(\omega t)$ and $\cos(3\omega t)$. This gives us two equations
\begin{eqnarray}
\omega^2k_1 + 0.75B^4+3 B^3+6 B^2+4 B+2&=&0 \nonumber \\ \label{eq:coeffeqns}
\omega^2k_2 + 0.5B^4+3B^3+3B^2+4B &=& 0 \hspace{0.75cm}(*)
\end{eqnarray}
where $k_1$ and $k_2$ are given below; 
\begin{eqnarray}
k_1&=&3.5625A^2B^2 + 7B^2C^2 + 2.125C^2B^4 -\\ \nonumber
   & & 2BC^2+3.5C^2B^3-0.75A^2-C^2, \\
k_2&=&0.1875A^2B^2 + 1.25A^2B -0.25A^2 +\\ \nonumber
   & &6.5C^2B^3 - 3.5B^2C^2 + 2BC^2 + 0.25C^2B^4.\hspace{0.25cm}(*)
\end{eqnarray}
The third equation needed to determine $A$, $B$ and $\omega$ is given by the initial condition
\begin{equation}
(1+B)x_0-A=0 \label{eq:initconds}
\end{equation}
 where $x_0 \equiv x(t=0)$.

We now solve these equations for  $A$, $B$ and $\omega$
with the given values of $x_0$. The solutions, as well as all mathematical
manipulations presented in this paper unless otherwise mentioned, were
determined using the software package Maple~15. The solutions have
$\omega^2$ $>$ 0 and are presented in Table~\ref{ta:upsoln}.

\begin{table}
    \begin{tabular}{ |l l l l l| p{3cm} |}
    \hline
   & $x_0$ & $A$ & $B$ & $\omega$ \\ \hline
1  & 0.001 & 0.001000006 & 6.2497E-6 & 14.142  \\ 
2  & 0.01 &  0.01000621694 & 0.0006216935078 & 14.089 \\ 
3  & 0.02 & 0.02004895506 & 0.002447752803 & 13.935 \\ 
4  & 0.03 & 0.03016097774 & 0.005365924678 & 13.689 \\ 
5  & 0.04 & 0.04036817727 & 0.00920443171 & 13.368 \\ 
6  & 0.05 & 0.05068761968 & 0.01375239358 & 12.989 \\ 
7  & 0.06 & 0.06112700808 & 0.01878346808 & 12.569 \\ 
8  & 0.07 & 0.07168549791 & 0.02407854158* & 12.125 \\ 
   & & (0.024074) &   \\
9  &0.08 & 0.08235549842 & 0.02944373* & 11.668 \\ 
   & & (0.029436) &  \\
10 &0.09 & 0.09312493232* & 0.03472147* & 11.211 \\ 
   & & (0.093124) & (0.034708) &  \\
11 & 0.1 & 0.1039794453* & 0.039794453* & 10.761 \\ 
   & & (0.103977) & (0.039775) &  \\
12 & 0.125 & 0.13139283* & 0.05114264* & 9.696 \\ 
   & & (0.131388) & (0.051102) &  \\
13 &0.15 & 0.1590498684* & 0.060332456* & 8.747 \\ 
   & & (0.159040) & (0.060266) &  \\
14 &0.175 & 0.1868185967* & 0.067534838* & 7.921 \\ 
   & & (0.186803) & (0.067443) &  \\
15 &0.2 & 0.2146246114* & 0.073123057* & 7.210* \\ 
   & & (0.214602) & (0.073008) & (7.209) \\
16 &0.225 & 0.2424301778* & 0.077467457* & 6.597 \\ 
  & & (0.242400) & (0.077332) & \\
17&0.25 & 0.27021796* & 0.08087184* & 6.068 \\
  & & (0.270180) & (0.080719) & \\
    \hline
    \end{tabular}
\caption{A comparison of our results with those of Contopolous and Seimenis (1990). Where a  corrected value appears, the original value appears in brackets
  below it.}
\label{ta:upsoln}
\end{table}

We note that for motion solely in the $x$-direction the value of $U$
is irrelevant, and it appears neither in Eq.~\ref{eq:coeffeqns}
nor in the initial conditions.

\begin{figure*}
\centerline{\psfig{figure=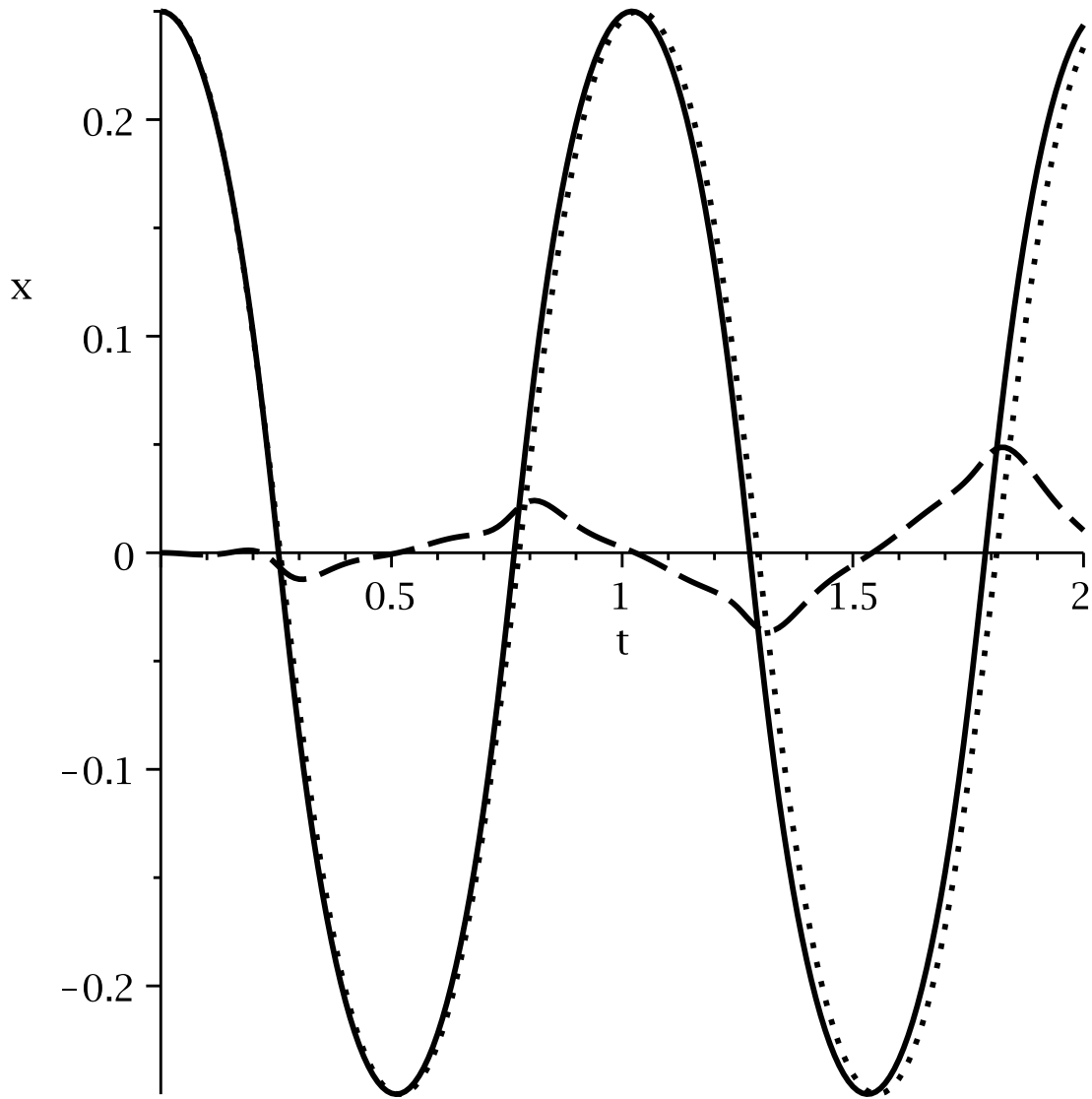,width=6in}}
\caption{A typical solution showing $x$ versus time for the unperturbed solution based on the values of line~17 in Table~\ref{ta:upsoln}. The solid line is the numerical solution from Maple 15, the dotted line is our approximation, and the dashed line is the difference between the two.}
\label{fi:soln1}
\end{figure*}

An example of an unperturbed solution is displayed in
Figure~\ref{fi:soln1} with values corresponding to line 17 in
Table~\ref{ta:upsoln}.

\section{Perturbed Solutions} \label{psection}

Purely radial orbits such as those of Section~\ref{upsection} are
unlikely in practice. Here the search is for solutions to the motion where
the $y$-component of the motion is close to the unperturbed motion
discussed previously. In this case, $M$ is no longer identically zero and
we look for solutions of the form
\begin{eqnarray}
M = M_0 + \delta M; \indent D = D_0 + \delta D; \indent N = N_0 + \delta N
\end{eqnarray}
where the subscript 0 indicates the unperturbed solution. The next
step is to solve the differential equation, introduced as Eq.~8 in \cite{consei90}
\begin{eqnarray}
(U^2N^2D^2+C^2U^2D^4 )\delta M''- \nonumber \\ \ (2U^2N^2 DD'+2C^2U^2D^3 D' )\delta M'  + \nonumber \\  (2U^2N^2D'^2-U^2N^2 DD''-C^2U^2D^3 D'' +\nonumber\\  2C^2U^2D^2 D'^2+2D^4 )\delta M=0. \label{eq:peqn}
\end{eqnarray}
The proposed solutions from Eq.~\ref{eq:propsoln} are
substituted into Eq.~\ref{eq:peqn} and solutions of the form
\begin{equation}
\delta M=\sum_{k=-\infty}^{\infty} C_k \cos \left( \left(\nu+k \right)\omega t \right) \label{eq:propsoln2}
\end{equation}
are searched for, where $\nu$ is a constant. CS determined from
\cite{flo83}'s work that values outside the range $0 \le \nu \le 1/2$
are unstable, and thus that $\nu(x_0)=0$ and $\nu(x_0)=1/2$ bracket
the stable region. They found no solution in the case of $\nu=0$, but
solutions do exist for the case $\nu=1/2$,  discussed below.

In order to get a finite number of non-trivial solutions,
Eq.~\ref{eq:propsoln2} must be truncated after a finite number of
terms. Following CS we take
\begin{equation}
\delta M = \sum_{k=-3}^{2} C_k \cos \left( \left(k+\frac{1}{2} \right)\omega t\right) \label{eq:propsoln2trunc}
\end{equation}
which leaves us with six values of $C_k$ to be determined.

The main goal here is to solve for the six constants $C_k$. In order
to do this, we substitute Eq.~\ref{eq:propsoln2trunc} and its
derivatives into Eq.~\ref{eq:peqn}, as well as the corresponding
substitutions for $N$ and $D$.  From this point on, we diverge from
the treatment of CS, as here we have used different methods to find
this equation's solutions. Here we used Maple~15 and Mathematica~8 as
tools for equation solving.

1) Using Maple's COMBINE function, Eq.~\ref{eq:peqn} was
solved for one value of $C_k$. The solution revealed many cosine terms with
different frequencies, and some terms that were fully independent of
the cosine.

2) The three lowest frequencies of cosine (including the independent
terms when present) were factored out of each individual $C_k$
term. The terms relating to a single frequency were collected,
yielding three separate equations. The cosine was then factored out,
and the remainder of the equations set equal to zero.

3) Steps 1 and 2 were repeated for each individual term of $C_k$.  The
result was 18 equations where there were 3 equations for each $C_k$ (each
of the three equations representing a different frequency of
cosine). Using these equations, a $6 \times 6$ matrix results, where rows 1, 3 and 5
represent equations with $k$ values of -3, -1, 1 and rows 2, 4, 6
represent equations with $k$ values of -2, 0, 2.
\begin{equation}
\left[ \begin {array}{cccccc} {\it S_{11}}&0&{\it S_{13}}&0&{\it S_{15}}&0\\ 
\noalign{\medskip}0&{\it S_{22}}&0&{\it S_{24}}&0&{\it S_{26}}
\\ \noalign{\medskip}{\it S_{31}}&0&{\it S_{33}}&0&{\it S_{35}}&0
\\ \noalign{\medskip}0&{\it S_{42}}&0&{\it S_{44}}&0&{\it S_{46}}
\\ \noalign{\medskip}{\it S_{51}}&0&{\it S_{53}}&0&{\it S_{55}}&0
\\ \noalign{\medskip}0&{\it S_{62}}&0&{\it S_{64}}&0&{\it S_{66}}\end {array}
 \right]  \label{eq:Smatrix}
\end{equation}
In other words, the columns are in increasing order from -3 to 2,
which demonstrates which equations contain which $k$ values.

We now solved the equations in sets of three for the values of
$C_k$. In essence, we constructed equations from the matrix ({\it
e.g.} $S_{11}+S_{13}+S_{15}=0$ and so forth down the rows). Rows 1, 3,
5 were used to solve for $C_{-3}$, $C_{-1}$ and $C_{1}$. Similarly,
rows 2, 4, 6 were used to solved for $C_{-2}$, $C_{0}$ and
$C_{2}$. The two homogeneous sets of three equations were transformed
to two non-homogeneous systems of order two with $C_{-1}$ and
$C_{0}$ set equal to unity for mathematical convenience. 

Once all six coefficients were determined, the values for $x(t)$ and
$y(t)$ could be determined for specific sets of values of $A$, $B$ and
$\omega$ from Table~\ref{ta:upsoln}. We then use the relations that
$u=\frac{1}{r}$, and that $r=\sqrt{x^2+y^2}$ to solve for $u$.

\begin{figure*}
\centerline{\psfig{figure=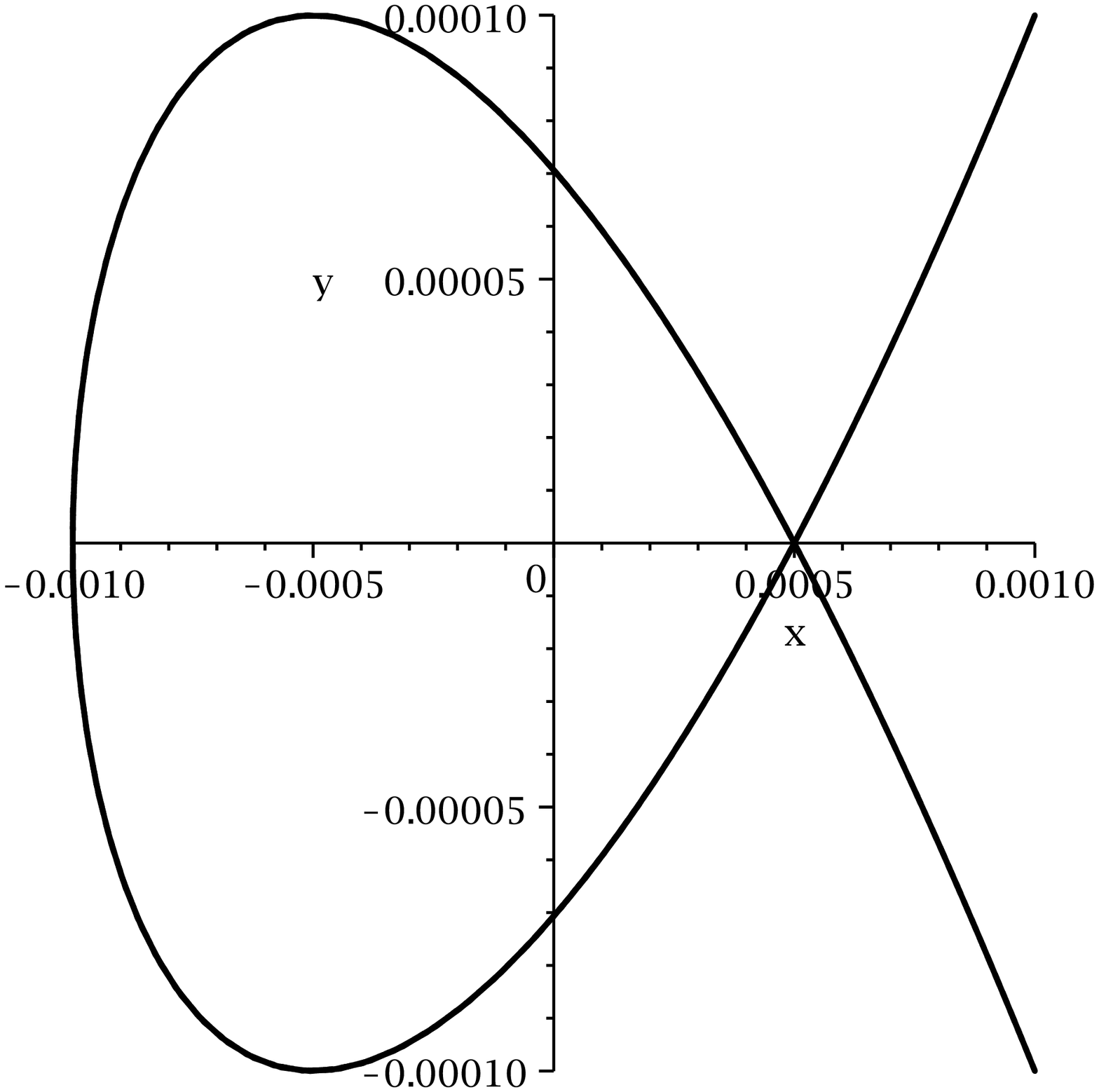,width=3in} \psfig{figure=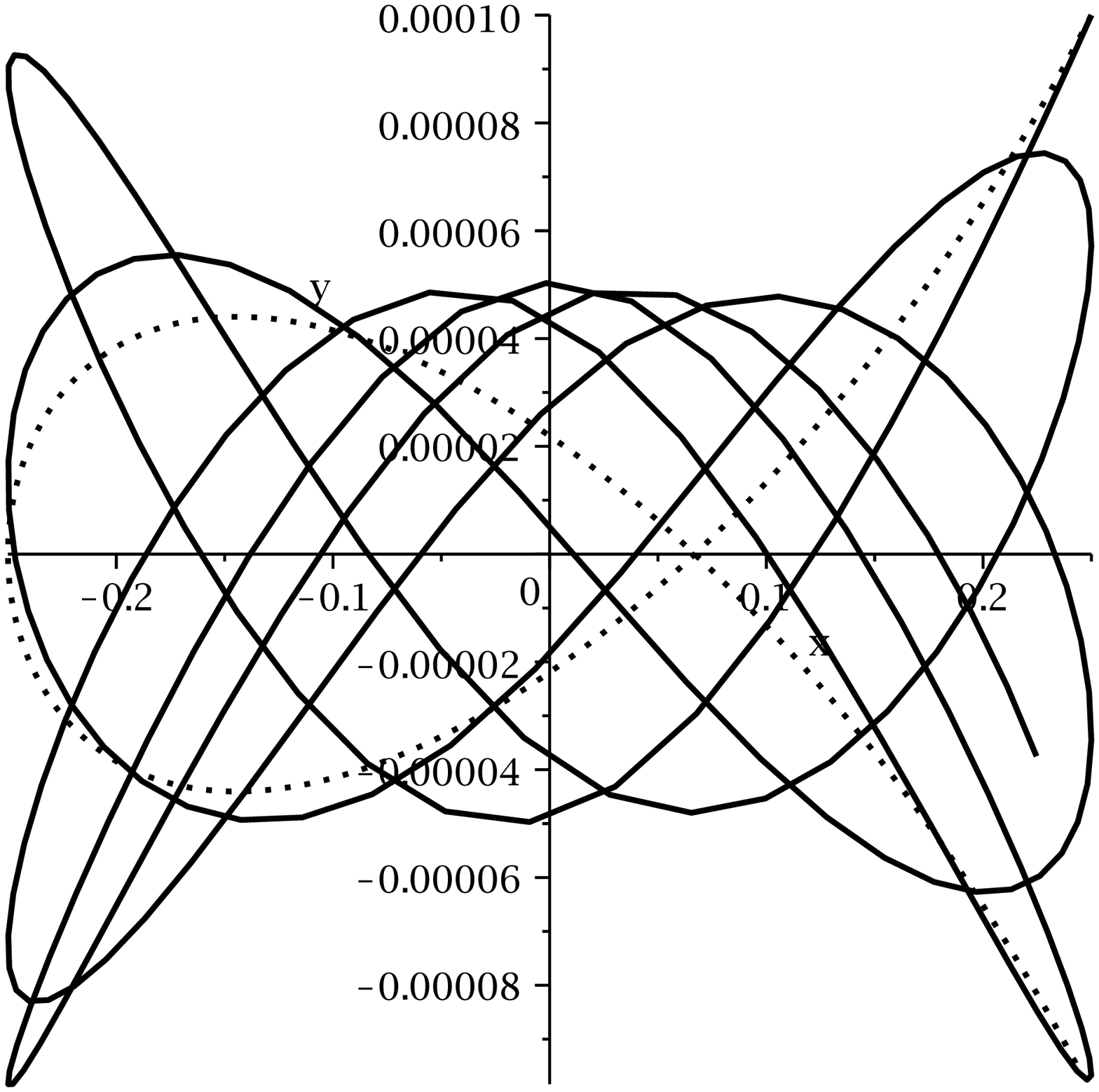,width=3in}}
\caption{Two solutions showing $x$ and $y$ as a parametric curve in $t$. The left panel shows the Prendergast and real solutions based on the values of line~1 in Table~\ref{ta:upsoln}. The two solutions plot on top of each other at this resolution and cannot be distinguished. The right panel shows a larger orbit, corresponding to the values of line~17 in Table~\ref{ta:upsoln}. The solid curve is the numerical solution from Maple 15 over a four cycles and the dotted line is the Prendergast approximation.}
\label{fi:soln2}
\end{figure*}

Figure~\ref{fi:soln2} shows two examples of a perturbed solution.
In the left panel, a parametric plot in $t$ of $x$ versus $y$ for
parameters $U=2/3$, $C$=0.1, and other values corresponding to line 1
in Table~\ref{ta:upsoln} is shown. Here the initial value of $y$ is
taken to be $10^{-4}$ to justify our assumption that it is small. The
two solutions are so similar that the graphs overplot each other and
cannot be distinguished. The right panel shows a much larger orbit
based on the parameters in line~17 of Table~\ref{ta:upsoln}. Here the
Prendergast solution does not contain enough frequency information to
completely reproduce the box orbit but captures some of the character of the
true solution, such as the $x$ and $y$ amplitudes and period.

\section{The Apsidal Angle and p-Ellipse Orbits} \label{pellsection}

In this section, we turn our interest to the approximate solution of
the radial orbital differential equation. We are
interested primarily in precession of the apsidal angle, and so
we will consider the problem now in terms of the anomaly $\theta$
rather than the time $t$. If one wished to determine the
relationship between these two variables, a Kepler-like
equation would need to be solved.

CS considered the $x$ and $y$ equations of motion, but here we
consider $u=1/\sqrt{x^2 +y^2} = 1/r$ with an eye to later using this
result to determine the apsidal angle for the purely radial
logarithmic potential. In this case, we examine the case where
$U \approx 1$ and $C \ll 1$.  We start with the radial orbital
differential equation, which is of the form,
\begin{equation} \label{eq:raddiff} 
\frac{d^2 u}{d \theta^2}
 +u=-\frac{1}{h^2 u^2 } f \left(\frac{1}{u} \right)
 \end{equation}
where $h$ is the angular momentum. We note that the force function
$f(\frac{1}{u} )$ is equal to $-dV/dr$, which can be obtained by
differentiating Eq.~\ref{eq:logpotl}.

The Prendergast Method works very well for the solutions of the
logarithmic potentials from Eq.~\ref{eq:logpotl} indicated in
Section~\ref{upsection} for the $x$ and $y$ equations of motion and
further elaborated in Section~\ref{psection}. However, the method does not seem to
be well suited for purely radial logarithmic potentials with different
initial conditions, and the solution does not agree with that obtained
by pure numerical integration of the orbital differential
equation. The p-ellipse approximate solutions of the orbital equation,
pioneered by \cite{str06}, is a much better way of not only deriving
an accurate approximate solution to order $e^2$ (where $e$ is the
orbital eccentricity), but also obtaining the values for the apsidal
precession to a high accuracy. We present a detailed analysis of the
orbital equation for the radial logarithmic potential with or without
the inclusion of the core scale length. In our analysis, the factor
$0 \le C \le1 $ gives a measure of the core scale length \cite[]{str06}.

For the case of large orbits, or negligible core size softening
length, the equation of motion is given by
\begin{equation} \label{eq:eom} 
 uu''+u^2=c 
\end{equation} 
where the parameter $c$, in the notation of Struck, depends both on
the constant scale mass $M^*$ and the core scale length $\varepsilon$
of the potential, the gravitational constant $G$, and the angular
momentum $h$. The above equation is similar to Eq.~\ref{eq:raddiff},
and of the form
 \begin{equation} \label{eq:eom2}
 u''+u=\frac{c}{u} 
 \end{equation} 
 Struck suggested an approximate solution of Eq.~\ref{eq:eom2}
\begin{equation} \label{eq:ellipse1}
u=\frac{1}{p} \left\{ 1+e \cos \left[ \left(1-b\right) \phi \right] \right\}^{1/2}
\end{equation}
 
Here, $p = a (1-e^2)$ is the semilatus rectum; $a$ is the semi major
axis ($e.~g.$ \cite{murder99,valyusmi05}), $e$ is the orbital eccentricity,
and $(1-b)$ is the factor associated with the precession
rate. Henceforth, for convenience, we set $k=1-b$ in our analysis, and
we will use $\theta$ instead of $\phi$ as was used by Struck. 

Struck finds that orbits obtained from a numerical integration of the
above differential equation look like precessing ellipses (p-ellipses)
and considers the approximate solution given in Eq.~\ref{eq:ellipse1}.

Substituting the solution of Eq.~\ref{eq:ellipse1} into the differential equation, we find that 
\begin{equation}
 -\frac{k^2}{2p^2} \left[-\frac{e^2-1}{2 \left(1+e \cos k \theta \right) } + \frac{1+e \cos k \theta} {2} \right] + \frac{1 + e \cos k \theta }{ 2 p^2} =c  
\end{equation}
\begin{equation} \label{eq:work1}
 \therefore  uu''  + u^2 =\frac{k^2}{2p^2} \frac{1 - e^2 +\frac{1}{p^2}(1-\frac{k^2}{4})(1+e\cos k \theta )}{2(1+e \cos k\theta )}=c
\end{equation}

The LHS of Eq.~\ref{eq:work1} simplifies to
 \begin{eqnarray}
uu''+u^2 &=& - \frac{1}{2p^2}  \left(1+e \cos k \theta \right) + \frac{1}{p^2} \left(1+e \cos k \theta \right) + \nonumber \\
 &&\frac{1}{2p^2} \left(1-e^2 \right) \frac{1}{1+e\cos k \theta } =c 
\end{eqnarray}
 Where $c_1=\frac{1}{2p^2} + \frac{1}{2p^2} =\frac{1}{p^2}$ and $k^2=
 (1-b_1)^2 =2$ are the first approximations to $k$ and
 $c$ \cite[]{str06}.
In a more accurate approximation to order $e^2$, we find that the constant terms reduce to
\begin{equation}  \label{eq:work2}
\frac{k^2}{4p^2} \left(1-e^2 \right) - \frac{k^2}{4p^2} + \frac{1}{p^2}=c 
\end{equation}

Comparing next, the terms involving $\cos k \theta$, we find that the coefficient of $\cos k \theta $ is given by 
\begin{equation} \label{eq:coeff1}
 - \frac{k^2}{4 p^2} e \left(1-e^2 \right) - \frac{k^2e}{4p^2} + \frac{e}{p^2} = 0 
\end{equation}
 
On simplification, one obtains 
\begin{equation} \label{eq:kseries}
k^2 = \frac{2}{1- \frac{1}{2}e} = 2 \left(1+ \frac{e}{2} + \frac{e^2}{4} \right) 
\end{equation}
For $e=0$, $k^2=2$ in accord with \cite{str06}.

In the case of non negligible core size, one has a similar though modified differential equation of the form
\begin{equation}
(uu''+u^2)(1+u^2)=c 
\end{equation}
The solution given in Eq.~\ref{eq:ellipse1} upon substitution into the above differential equation leads to the expression
\begin{eqnarray}
 &&\left[ -\frac{k^2}{4p^2} \left(1+e \cos k\theta \right) + 1 + \frac{e \cos k \theta}{p^2} + \frac{k^2}{4p^2} \frac{\left(1-e^2 \right)}{1+e \cos k \theta} \right] \times \nonumber \\
 && \left(1+\frac{1+e\cos k \theta}{p^2} \right)=c
\end{eqnarray}
which, upon comparison of terms independent of $\cos k \theta$,  simplifies to
\begin{eqnarray}
&& \frac{1}{p^2} \left(1-\frac{k^2 e^2}{4} \right) \left(1+ \frac{1}{p^2} \right) + \frac{k^2}{4p^2} \left(1-e^2 \right) \left(1+\frac{1}{p^2} \right) \left(- \frac{e^2}{2} \right) + \nonumber \\
&&\frac{k^2}{4p^2} \left(1-e^2 \right) \left(-\frac{e^2}{2p^2} \right)=c
\end{eqnarray}

Comparing coefficients of $ \cos k \theta$, we get the more general dependence of $k^2$.
\begin{equation}
 k^2 = 2 \left[ 1 + \frac{\frac{e^2}{2} + \frac{1}{p^2}}{1+ \frac{1}{p^2} - \frac{e^2}{2}} \right] 
\end{equation}
 
If terms of order $e^2$ are ignored, 
\begin{equation}
 k^2 = 1 - b_1  =2 \left( \frac{1+ \frac{2}{p^2}}{1 + \frac{1}{p^2}} \right)
\end{equation}
where $b_1$ is the first approximation of the precession factor.

Furthermore, to order $e^4$
\begin{eqnarray}
 c &=& \frac{1}{p^2} \left(1 + \frac{1}{p^2} \right) - \frac{k^2 e^2}{4p^2} \left(\frac{3}{2} + \frac{1}{p^2} + \frac{2}{p^2} \right) + \nonumber \\
& &\frac{k^2 e^4}{8p^2} \left(1+\frac{2}{p^2} \right)  
\end{eqnarray}
 in agreement with Struck.

 It is interesting to note that the orbital differential equation
 associated with apsidal precession is a special case of the
 generalized Burgers partial differential equations (GBE) and seems to
 characterize these equations similar to the way that Painleve
 equations represent the Korteweg-de Vries type of
 equations \cite[]{sac91}. This variety of equations can be expressed as
 Eqs.~\ref{eq:pl1} and \ref{eq:pl2} where $f(x)$ and $g(x)$ are
 sufficiently smooth arbitrary functions, $a,e$ and $c$ are real
 constants, and the solutions of $y$ are Euler-Painleve
 transcendents \cite[]{kam43}.
\begin{equation} \label{eq:pl1}
 yy'' + ay'^2 + f(x)yy' + g(x) y^2 + ey'+c=0
\end{equation}

 In the case where $f(x)$ and $g(x)$ are constants, we have the Euler-Painleve equation
\begin{equation} \label{eq:pl2}
 yy'' + ay'^2 + byy' + cy^2 + dy^{1 -\alpha}=0
\end{equation}
 
For $\alpha =1, b=0$, the substitution $y=u^{1/2}$ leads to the differential equation
\begin{equation}
 -\frac{1}{4}\frac{u'^2}{u} + \frac{1}{2}u'' + cu+d +\frac{a}{4} \frac{u'^2}{u} = 0
\end{equation}
 For $a=1$ the $\frac{u'^2}{4u}$ terms cancel, and the following equation results.
\begin{equation}
 \frac{1}{2}u''+cu+d=0
\end{equation}

It is important to observe that the orbital differential equation does
not have the $ay'^2$ term in contrast to the GBE. This term contains
terms of order $e^2$ and the correction does not turn out to be
significant. Hence, the $p$ ellipse is a natural approximate solution
of the generalized Burgers equations (GBE) and is an Euler -- Painleve
transcendant \cite[]{kam43}.

 As a rough estimate of the mean error in neglecting the
 $\frac{u'^2}{4u}$ term, we evaluate the following integrals that
 occur in the evaluation of this term.
\begin{equation}
 -\frac{2k^2 e^2}{16p}\frac{1}{e^2} \left[ \int^\pi_0 \left(1+e \cos k \theta \right)^{\frac{1}{2}} d\theta -\int^{\pi}_0 (1+e \cos k \theta )^{-\frac{1}{2}} d\theta \right]
\end{equation}
 With the substitution $k\theta =x$ we have
\begin{eqnarray}
 I_1 &=&  \left[ \frac{1}{\pi} \frac{1}{k} \int^{k\pi}_0 (1+e\cos x)^{\frac{1}{2}} dx \right] \nonumber \\
  &=& \left[ 1+\frac{1}{2}e \frac{\sin k\pi}{k \pi} -\frac{1}{8k \pi} e^2 \left(\frac{k\pi}{2} +\frac{\sin 2k \pi}{4}\right) + \dots \right] 
\end{eqnarray}
 and
\begin{eqnarray}
I_2 &=& \frac{1}{\pi}\frac{1}{k} \int^{k \pi}_0 \frac{dx}{\left(1+e\cos x\right)^{\frac{1}{2}}} \nonumber \\
   &=& \left[1-\frac{1}{2}e \frac{\sin k\pi}{k\pi} + \frac{3}{16}e^2 + \frac{3}{16}e^2 \frac{\sin 2k\pi}{2k\pi} + \dots \right] 
\end{eqnarray}
Hence, we obtain for the difference of the two integrals, 
\begin{equation}
I_1-I_2 \approx \left[e\frac{\sin k\pi}{k\pi} - \frac{e^2}{4} - \frac{2e^2}{16} \frac{\sin 2 k\pi}{k\pi} + \dots \right]
\end{equation}
 
An approximate mean error (M.E.) due to the presence of the term  $\frac{u'^2}{4u}$  is
\begin{equation}
{\rm M.E.} = \left|- \frac{2k^2}{16\pi} \frac{1}{p} \left[ - \frac{0.9 e}{k\pi} - \frac{e^2}{4} + \frac{e^2}{16} + \dots \right]\right|
\end{equation}
Recalling $p = a(1-e^2)$ and taking $k\approx 1.45$ and $e\sim0.9$
\begin{equation}
{\rm M.E.} = \frac{2.9}{16\pi} \frac{1}{0.19 a} \left|- \frac{0.81}{4} - \frac{2.43}{16} \right| 
\end{equation}
We find that M.E. is $\sim 1 \%$ for $a=1$, $e=0.3$ ($k=1.79$); M.E. increases with
higer $e$ and decreases with larger $a$.

Interestingly, when $k \theta =\pi$ or $k\theta =0$,
\begin{equation}
 \frac{u'^2}{4u} =0 
\end{equation}
 showing that this correction term does not contribute to these angles, as shown below.
\begin{eqnarray}
  \frac{-(1-e^2)}{\left(1-e\right)\left(1-e\right)^{\frac{1}{2}}} + \frac{1+e}{\left(1-e\right)^{\frac{1}{2}}}  &=& -(1-e^2)+(1+e^2) \nonumber \\
&=&0 
\end{eqnarray}
 
 Next, we calculate the apsidal angle for the orbits in a logarithmic
 potential. The apsidal angle is the angle at the force centre between
 the smallest and largest apses, that is, between pericenter and
 apocenter. Hence, the behaviour of the logarithmic potential is
 similar to that of the $n>2$ power law potentials. Thus, there
 will always be a single minimum regardless of the value of the
 constant $c$. As $c$ increases the location of the minimum simply
 shifts to larger $x$ values.

 Only bound orbits are possible for this potential. As
 $x \rightarrow \infty$, $V(x)$ also approaches infinity due to the
 $\ln (x)$ term, so there is always an inner and an outer turning
 point no matter how large the total energy of the system. Stable
 circular orbits are possible at the minimum of the effective potential.

 The approximate $p$-ellipse orbits are, on first appearance, only
 good to first order in $e$. However, Struck, in his thorough
 analysis, has shown that the orbital fits are excellent over several
 orbital periods. In fact, the value of $k=1-b$ which more accurately
 depends on $e$, is still fairly close to the more exact value; as
 partly due to the slow variation of $c$ with $e$. The apsidal angle
 has been calculated for various values of $e$ and is shown in
 Table~\ref{ta:eccentricity}.

\begin{table}
\begin{center}
 \begin{tabular}{cccc}  \hline
 $a$ & $e$  & $k$     & $\theta$ (rad)  \\ \hline
 0.75&  0.1 & 1.81363 & 1.73222    \\
 0.75&  0.5 & 1.87598 & 1.67464    \\
 0.75&  0.9 & 1.99002 & 1.57867    \\
 1   &  0.1 & 1.73495 & 1.81077    \\
 1   &  0.5 & 1.81108 & 1.73465    \\
 1   &  0.9 & 1.98250 & 1.58466    \\
 1.5 &  0.1 & 1.62362 & 1.93493    \\
 1.5 &  0.5 & 1.78858 & 1.75647    \\
 1.5 &  0.9 & 2.17205 & 1.44637    \\
 6   &  0.1 & 1.43715 & 2.18599    \\
 6   &  0.5 & 1.53520 & 2.04637    \\
 6   &  0.9 & 1.98441 & 1.58314    \\ \hline
    \end{tabular}
\caption{Some values of the apsidal angle from the p-ellipse prescription ($C=0$) with varying eccentricity.}
\label{ta:eccentricity}
\end{center}
\end{table}

We now calculate the apsidal angle by using the Lambert $W$ function,
a function that is creating a renaissance in solving many interesting
problems involving roots and limits of integration, as well as others.

We begin by defining the energy $E$ of an orbit through the summation
of its kinetic and potential energies:
 \begin{equation} \label{eq:energy}
 E=\frac{1}{2} \left[\left(\frac{dr}{dt} \right)^2 +r^2 \left(\frac{d\theta}{dt} \right)^2 \right]+V(r) 
 \end{equation} 
where $r^2 \frac{d\theta }{dt} =h$ is the angular momentum, and
$\frac{dr}{d\theta } $ can be broken into
$\frac{dr}{du} \frac{du}{d\theta } $.

Our main goal is to solve for $\frac{du}{d\theta } $ as this will
provide us with an integrable function from which we can ultimately
obtain a value for $\theta $.

Working in the regime where $C \ll 1$ and $U \approx 1$, $V(r)$ can be
simplified further and we obtain the following
\begin{equation} \label{eq:dthetadu}
\frac{d\theta }{du} =\frac{h}{\sqrt{2E+2\ln u^{2} -u^{2} A} }  
\end{equation} 
where $A=h^2 +2C^{2}$ and $E=0.5+\ln r_{c} $, where $r_{c}$ is the
radius of the ($c$ for 'circular') orbit. The value of $u_{c} =1/r_{c} $ is
taken here at values between 1 and 1.8, examining a range around the
nominal value ($E\approx 0$, $h=e^{-1/2} $) of $u_{c} =e^{1/2} \approx
1.648$.

Where $d\theta /du$ passes from positive to negative reveals the
location of the apses, thus the limits of integration of
Eq.~\ref{eq:eom2} are its corresponding roots. We can solve for these
roots by setting the denominator equal to zero, and manipulating it so
that it can become solvable using the Lambert $W$ function  \cite[]{valjefcor00}. We
start by reworking the denominator into the following form:
\begin{equation}
 \ln u^{2} -\frac{u^{2} }{2} A=-E \label{eq:Wfn1}
\end{equation} 
The roots are given by the expression
\begin{equation}  
u=\sqrt{\frac{-2W_{j} (-\frac{A}{2} e^{-E} )}{A} }
\end{equation}
where $W_{j}$ represents the Lambert $W$ function and $j$ represents
the chosen branch. We solve for the two roots by using the -1 and the
zero branches.

Having the apocenter $r_{M} $ and pericenter $r_{m} $ distances in
hand allows a determination of the orbit eccentricity through
\begin{equation}
\frac{r_{M} }{r_{m} } =\frac{1+e}{1-e} . 
\end{equation} 
We note that solutions with imaginary eccentricity, which have two
complex solutions which are conjugates of each other, would be
manifested by a plunging of the orbit into the force centre \cite[]{hag31,cha83}.

Integrating Eq.~\ref{eq:dthetadu} with the two roots as end points of
the integral yields an answer that represents the apsidal angle for
the particular orbit with a specific value of $u_{c}$.

\begin{figure*}
\centerline{\psfig{figure=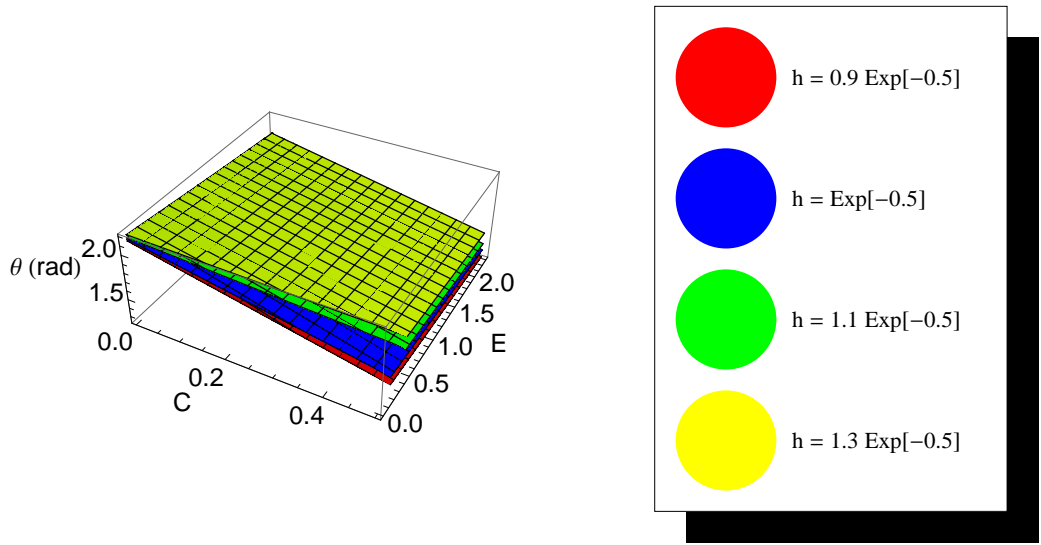,width=6in}}
\caption{The apsidal angle in radians as a function of the core radius  $C$ and the energy $E$. Four different angular momenta are presented in separate colours. Please see the text for more details.}
\label{fi:apsidalangle}
\end{figure*}

Figure~\ref{fi:apsidalangle} shows the apsidal angle calculated by this method, for different values of $C$, $E$ and $h$.

\begin{table}
\begin{center}
 \begin{tabular}{cccc}  \hline
 $u_c$ & \multicolumn{3}{c}{$\theta$} \\ \hline
    & Lambert $W$   & Numerical & Difference \\
    & approximation &           &            \\ \hline
 1.0 & 2.06310 & 2.06300 & 0.00010 \\ 
 1.1 & 2.07122 & 2.07116 & 0.00006 \\  
 1.2 & 2.07868 & 2.07862 & 0.00006 \\  
 1.3 & 2.08558 & 2.08554 & 0.00004 \\  
 1.4 & 2.09201 & 2.09200 & 0.00001 \\  
 1.5 & 2.09803 & 2.09797 & 0.00006 \\  
 1.6 & 2.10368 & 2.10361 & 0.00007 \\  
 1.7 & 2.10901 & 2.10896 & 0.00005 \\  
 1.8 & 2.11405 & 2.11397 & 0.00008 \\  \hline
    \end{tabular}
\caption{The apsidal angle as calculated for different values of $u_c = 1/r_c$.}
\label{ta:apsidal}
\end{center}
\end{table}
 
Table~\ref{ta:apsidal} shows how values of $u_{c}$ ranging from 1 to
1.8 yield similar apsidal angles with values near
$\frac{2\pi}{3}$. For comparison, \cite{toutre97} use the epicyclic
approximation for near-circular orbits to determine that their
$g(\alpha ,y)$ (which is twice the apsidal angle as defined here)
equals $2\pi /\sqrt{2} =4.44428=2\times 2.22144$, a value close to the
one arrived at here. However, a comparison with the
numerically-derived result, also listed in Table~\ref{ta:apsidal} shows that the
method proposed here is much more accurate: the two differ only in the
fifth decimal place. As a comparison, we also show the apsidal angle
for various values of small $e$ using the p-ellipse
approximation in the column labelled 'Numerical'.

 From Eq.~\ref{eq:kseries}, we have 
\begin{equation}
k=\sqrt{2}{{{\left(1-\frac{e}{2}\right)}}}^{{-1}/{2}}\ \sim \ \sqrt{2}{{\left(1+\frac{e}{4}+\dots \right)}}
\end{equation}
 The apsidal angle is given by
\begin{equation}
\frac{\pi }{k}=\frac{\pi }{\sqrt{2}}{\left(1-\frac{e}{2}\right)}^{{1}/{2}}\ \sim \ \frac{\pi }{\sqrt{2}}\left(1-\frac{e}{4}-\frac{1}{32}e^2\dots \right)
\end{equation}

It is of interest to note that the roots can be found without any
approximation for $C$ in terms of the polylog function. For arbitrary
$C$, one obtains from Eq.~\ref{eq:energy} an exact expression
\begin{equation} \label{eq:Cexpr1}
 -\frac{1}{C^{2} u^{2} } (\ln C^{2} -E)-\frac{1}{C^{2} u^{2} } \ln \left(1+\frac{1}{C^{2} u^{2} } \right)=\frac{h^{2} }{2C^{2} } 
\end{equation} 
for finding the roots of $u^{2}$. Now if we define $k\equiv \ln C^{2} -E$ and $x\equiv -\frac{1}{C^{2} u^{2} } $, Eq.~\ref{eq:Cexpr1} reduces to 
\begin{eqnarray}
kx+x\ln (1-x) &=&\frac{h^{2} }{2C^{2} } \label{eq:Wfn2}\\
k+\ln (1-x) &=&\frac{h^{2} }{2C^{2} } \frac{1}{x} \\
k-Li_{1} (x)&=&\frac{h^{2} }{2C^{2} } \frac{1}{x} \label{eq:polylog1} 
\end{eqnarray} 
Here Eq.~\ref{eq:polylog1} is the functional equation of the Polylogarithm $Li_{1} (x)=-\ln (1-x)$ \cite[]{lew81} and
\begin{equation}
x=-\frac{1}{C^{2} u^{2} } =-\frac{r^{2} }{C^{2} }
\end{equation}

 \section{Gravitational Lensing } \label{gravlensing}

The use of the Lambert $W$ and the Polylogarithm functions to find the
roots of equations such as Eq.~\ref{eq:Wfn1} and \ref{eq:Wfn2} may
have wider applicability. For example, we can use a similar approach
to compute the deflection of a light ray by a logarithmic potential,
useful in the context of gravitational
lensing \cite[]{cow83,sch90,bluschmor10}. \cite{zwi37} suggested that
extragalactic nebulae offer a much better chance than stars for the
observation of gravitational lens effects. Zwicky's idea was that some
of the massive and more concentrated nebulae may be expected to
deflect light by as much as half a minute of arc. Nebulae, in contrast
to stars, possess apparent dimensions which are resolvable to very
great distances. Zwicky was following up on the work of \cite{ein36}
on stars acting as a gravitational lens. According to Zwicky,
observations on the deflection of light around nebulae may provide the
most direct determination of nebular
masses \cite[]{smi36}. \cite{zwi37} estimated the probability of
detecting nebular galaxies which act as gravitational lenses and
pointed out the possibility of ring shaped images, flux amplification
and understanding the large scale structure of the universe. The
lensing equation can be generalized to three dimensions and
cosmological distances by correction of the redshift related
distance \cite[]{schehlfal92}.

For arbitrary $K$ one has the following expression to determine the
roots in the case of light deflection for a logarithmic potential,
\begin{equation}
V\left(r\right)= K \ln \left(r^2+C^2\right)
\end{equation}
where $K$ is a dimensionless constant and $r=\frac{1}{u}$.

For light deflection in the logarithmic potential considered, the differential equation for the given logarithmic potential is of the form
\begin{equation}
\frac{d^2u}{d\theta^2}+\ u= K\left(\frac{C^2u}{1+C^2u^2}-\frac{1}{u}\right)=\ \frac{-K}{u(1+C^2u^2)} 
\end{equation}
The DE for small values of $C \ll 1$, reduces to
\begin{equation}
\therefore \frac{d^2u}{d\theta^2} +u - KC^2 u=\frac{-K}{u}
\end{equation}
 
In the relativistic formulation \cite[]{har03} the differential equation for light deflection is
\begin{equation}
{\left(\frac{d\theta }{du}\right)}^2=\frac{1}{\frac{1}{b^2}-u^2+2Mu^3}
\end{equation}
\begin{equation}
\therefore u''+u=3Mu^2 
\end{equation}
where $b$ is the impact parameter.

 Assuming the photon is a non-relativistic particle that travels at
 speed $c$ and it is far from all sources of gravitational
 attraction \cite[]{har03}, we can determine the light deflection
 $\Delta \theta $ produced by a logarithmic potential as
 \begin{equation}
 \Delta \theta =2\int _{0}^{u_{1} } \frac{du}{\sqrt{1-u^{2} +K\ln \left(\frac{1}{u^{2} } +C^{2} \right)} }  
 \end{equation} 

 Solving for the roots of the denominator, one obtains 
\begin{equation}
1-u^2+ K \left[\ln \left(1+C^2u^2\right)- \ln u^2 \right]=0
\end{equation}

 Solving for $K \ll 1$ by use of the Lambert $W$ function, we get 
\begin{equation}
u^2 = \frac{1}{\left(\frac{1}{K}- C^2\right)} W_j\left\{\left(\frac{1}{K}-C^2\right)e^{\frac{1}{K}}\right\}
\end{equation}
 where $j$ denotes the branch of the Lambert $W$ function.

\begin{figure*}
\centerline{\psfig{figure=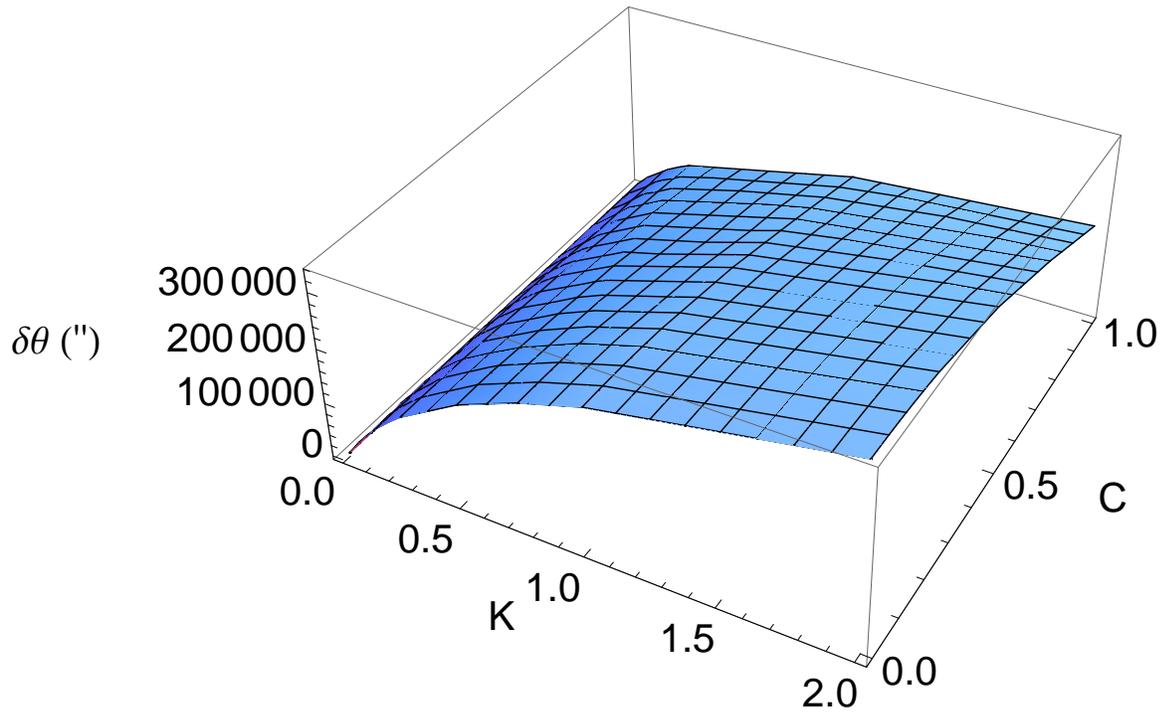,width=6in}}
\caption{The deflection angle in seconds of arc as a function of $C$ and $K$. Please see the text for more details.}
\label{fi:deflectionangle}
\end{figure*}

The deflection angle is related to the Einstein angle $\theta_E $
\cite[]{har03} which sets the characteristic angular scale for
gravitational lensing phenomena. Gravitational lensing can be used to
detect mass or energy in the universe, whether visible or
not. Table~\ref{ta:deflectionangle} show the deflection angle for a
range of values of $K$ and $C$. Small $K$ values produce small
deflections, while smaller values of $C$ produce larger ones, though
with a weaker dependence. Figure~\ref{fi:deflectionangle} shows the
variation graphically.

\begin{table}
\begin{center}
\begin{tabular}{ccc}
 $K$       & $C$    &  $\Delta \theta$ \\\hline
           & 1.0    &  0.00444\\
 $10^{-8}$  & 0.5    &  0.00703\\
           & 0.0005 &  0.00491\\ \hline
           & 1.0    &  2256.33\\
 0.005     & 0.5    &  2861.51\\
           & 0.0005 &  3202.15\\ \hline
           & 1.0    &  69 637.4\\
 0.25      & 0.5    &  95 986.9\\
           & 0.0005 & 109 909\\ \hline
           & 1.0    & 104 102 \\
 0.5       & 0.5    & 150 674 \\
           & 0.0005 & 173 731 \\ \hline
           & 1.0    & 142 042 \\
 1         & 0.5    & 218 327 \\
           & 0.0005 & 251 802 \\ \hline
           & 1.0    & 178 640 \\
 2         & 0.5    & 292 967 \\
           & 0.0005 & 334 185 \\    \hline
\end{tabular}
\caption{The deflection angle $\Delta \theta$ (in seconds of arc) as a function of $k$ and $C$.}
\label{ta:deflectionangle}
\end{center}
\end{table}

Analogous calculations can be done for time delay in light signals due
to lensing galaxies \cite[]{oharuf94}. Intensity fluctuations
caused by lumpy dark matter may provide direct observational existence
for it. It is worth noting that the entire analysis can not only be
done for the purely radial logarithmic potential, but also for the
Eq.~\ref{eq:logpotl} of the logarithmic potential, by considering the
$x$ and $y$ components separately as was done for the gravitational
potential \cite[]{boukan75}.

 \section{Conclusions} \label{conclusions}

We have revisited and expanded the work of \cite{consei90} on orbits
within a logarithmic potential. We did a comprehensive review of the
Prendergast method used by CS. We performed an analytic and numerical study
of the matrix: $C_{-3} $, $C_{-2} $, $C_{-1} $, $C_{0} $, $C_{1} $,
$C_{2} $ that resulted in eighteen equations for the six coefficients
for the orbital Fourier type series solution for values of $U$ ranging
from 0.1 to 1 that gave the unperturbed as well as perturbed solutions
with better precision. The apsidal angle for the case of galactic
orbits for a planar scale-free spherical logarithmic potential was
obtained from the p-ellipse solution of the orbital differential
equation and also the Lambert $W$.  Both the Lambert $W$ and the
Polylogarithm functions may have applications in problems involving
exponential and/or logarithmic potentials such as gravitational
lensing.

The Prendergast method, although not used as widely as others, has
been quite useful in our analysis in Sections 2 and 3., and is likely
to prove useful in the study of many types of galactic potentials.

Gravitational lensing can be used to detect mass in the universe,
whether dark or visible \cite[]{har03,nar10}. In general relativity,
all energy curves spacetime, and a constant vacuum energy produces a
detectable curvature. Gravity may prove a useful tool for detecting
and studying dark energy.  The lensing due to the gravitational field
of a black hole of background stars and galaxies \cite[]{tho94} can be
significant and the effects of a logarithmic potential warrant further
study in this connection.
 
\bibliographystyle{mn2e}
\bibliography{Wiegert}

\section{Acknowledgements}
 We gratefully acknowledge discussions with Dr. Seimenis during our
 work. We thank Curt Struck (Iowa State University) for sending
 earlier work on p-ellipse orbits. We also thank the anonymous
 referee for a stimulating review of our manuscript. SRV gratefully
 acknowledges research funding from King's University College at the
 University of Western Ontario. This work was supported in part by the
 Natural Sciences and Engineering Research Council of Canada (NSERC).

\begin{appendix}

\section{Perturbed equations}

The following are the equations for Equation~\ref{eq:Smatrix}
\begin{eqnarray}
 {\it S_{11}}={\it E_6}-{U}^{2}{\it E_1} \left( \nu-3 \right) ^{2}{\omega}^{
2}; \nonumber \\
{\it S_{13}}= 0.5{\it E_7}- 0.5{U}^{2}{\it E_2}{\omega}^{2} \left( \nu-
1 \right) ^{2}- 0.5{U}^{2}{\it E_4}\omega \left( \nu-1 \right); \nonumber \\
{\it S_{15}}= 0.5{\it E_8}- 0.5{U}^{2}{\it E_3}{\omega}^{2} \left( \nu+
1 \right) ^{2}- 0.5{U}^{2}{\it E_5}\omega \left( \nu+1 \right); \nonumber \\
{\it S_{22}}={\it E_6}-{U}^{2}{\it E_1} \left( \nu-2 \right) ^{2}{\omega}^{
2}; \nonumber \\
{\it S_{24}}= 0.5{\it E_7}- 0.5{U}^{2}{\it E_2}{\omega}^{2}{\nu}^{2}-
 0.5{U}^{2}{\it E_4}\omega\nu; \nonumber \\
{\it S_{26}}= 0.5{\it E_8}- 0.5{U}^{2}{\it E_3}{\omega}^{2} \left( \nu+
2 \right) ^{2}- 0.5{U}^{2}{\it E_5}\omega \left( \nu+2 \right); \nonumber \\
{\it S_{31}}= 0.5{\it E_7}- 0.5{U}^{2}{\it E_2}{\omega}^{2} \left( \nu-
3 \right) ^{2}+ 0.5{U}^{2}{\it E_4}\omega \left( \nu-3 \right); \nonumber \\
{\it S_{33}}={\it E_6}-{U}^{2}{\it E_1} \left( \nu-1 \right) ^{2}{\omega}^{
2}; \nonumber \\
{\it S_{35}}= 0.5{\it E_7}- 0.5{U}^{2}{\it E_2}{\omega}^{2} \left( \nu+
1 \right) ^{2}- 0.5{U}^{2}{\it E_4}\omega \left( \nu+1 \right); \nonumber \\
{\it S_{42}}= 0.5{\it E_7}- 0.5{U}^{2}{\it E_2}{\omega}^{2} \left( \nu-
2 \right) ^{2}+ 0.5{U}^{2}{\it E_4}\omega \left( \nu-2 \right); \nonumber \\
{\it S_{44}}={\it E_6}-{U}^{2}{\it E_1}{\omega}^{2}{\nu}^{2}; \nonumber \\
{\it S_{46}}= 0.5{\it E_7}- 0.5{U}^{2}{\it E_2}{\omega}^{2} \left( \nu+
2 \right) ^{2}- 0.5{U}^{2}{\it E_4}\omega \left( \nu+2 \right); \nonumber \\
{\it S_{51}}= 0.5{\it E_8}- 0.5{U}^{2}{\it E_3}{\omega}^{2} \left( \nu-
3 \right) ^{2}+ 0.5{U}^{2}{\it E_5}\omega \left( \nu-3 \right); \nonumber \\
{\it S_{53}}= 0.5{\it E_7}- 0.5{U}^{2}{\it E_2}{\omega}^{2} \left( \nu-
1 \right) ^{2}+ 0.5{U}^{2}{\it E_4}\omega \left( \nu-1 \right); \nonumber \\
{\it S_{55}}={\it E_6}-{U}^{2}{\it E_1} \left( \nu+1 \right) ^{2}{\omega}^{
2}; \nonumber \\
{\it S_{62}}= 0.5{\it E_8}- 0.5{U}^{2}{\it E_3}{\omega}^{2} \left( \nu-
2 \right) ^{2}+ 0.5{U}^{2}{\it E_5}\omega \left( \nu-2 \right); \nonumber \\
{\it S_{64}}= 0.5{\it E_7}- 0.5{U}^{2}{\it E_2}{\omega}^{2}{\nu}^{2}+
 0.5{U}^{2}{\it E_4}\omega\nu; \nonumber \\
{\it S_{66}}={\it E_6}-{U}^{2}{\it E_1} \left( \nu+2 \right) ^{2}{\omega}^{
2}. \nonumber 
\end{eqnarray}

In $S_{22}$ the bracketed term $(\nu-2)$ has been corrected from the original form with $(\nu-1)$ in CS.

The corresponding $E$ values are
\begin{eqnarray}
{\it E_1}={C}^{2} \left(  0.375{B}^{4}+3{B}^{2}+1 \right) + 0.5{A
}^{2} \left(  0.5{B}^{2}+B+1 \right); \nonumber \\
{\it E_2}={C}^{2} \left( 3{B}^{3}+4B \right) +{A}^{2} \left(  0.375
{B}^{2}+B+ 0.5 \right); \nonumber \\
{\it E_3}={C}^{2} \left(  0.5{B}^{4}+3{B}^{2} \right) + 0.5{A}^{2
} \left(  0.5{B}^{2}+B \right); \nonumber \\
{\it E_4}=2\omega B \left( {A}^{2}+2{C}^{2}+{C}^{2}{B}^{2}
 \right) +\omega{B}^{2} \left(  0.5{A}^{2}+2{C}^{2}B \right) -
\omega{B}^{3}{C}^{2}; \nonumber \\
{\it E_5}=\omega{B}^{2} \left( {A}^{2}+2{C}^{2}+{C}^{2}{B}^{2}
 \right) +\omega B \left( {A}^{2}+4{C}^{2}B \right); \nonumber \\
{\it E_6}={U}^{2}{\omega}^{2}B \left( {A}^{2}+3{A}^{2}B+10{C}^{2}B+
 2.5{C}^{2}{B}^{3} \right) + 0.75{B}^{4}+6{B}^{2}+2; \nonumber \\
{\it E_7}={U}^{2}{\omega}^{2}B \left( 2{A}^{2}+ 2.5{A}^{2}B+4{C}^
{2}+13{C}^{2}{B}^{2} \right) +6{B}^{3}+8B; \nonumber \\
{\it E_8}={U}^{2}{\omega}^{2}B \left( {A}^{2}-{A}^{2}B+2{C}^{2}B+2{
C}^{2}{B}^{3} \right) +{B}^{4}+6{B}^{2}. \nonumber 
\end{eqnarray} 
\label{lastpage}
In $E_4$ the term with $0.5A^2$ has been corrected from its original form of $0.5A$ in CS.

\end{appendix}

\newpage

\newpage

\newpage

\newpage


\end{document}